\documentclass{PoS}

\usepackage{color}

\def\Comment#1{}
\newcommand{\kw}{\kappa_{W}}

\newcommand{\bean}{\begin{eqnarray*}}
\newcommand{\eean}{\end{eqnarray*}}

\newcommand{\gapproxeq}{\lower
.7ex\hbox{$\;\stackrel{\textstyle >}{\sim}\;$}}
\newcommand{\lapproxeq}{\lower
.7ex\hbox{$\;\stackrel{\textstyle <}{\sim}\;$}}

\newcommand\lsim{\mathrel{\rlap{\lower4pt\hbox{\hskip1pt$\sim$}}
    \raise1pt\hbox{$<$}}}
\newcommand\gsim{\mathrel{\rlap{\lower4pt\hbox{\hskip1pt$\sim$}}
    \raise1pt\hbox{$>$}}}
\newcommand{\ba}{\begin{array}}
\newcommand{\ea}{\end{array}}
\newcommand{\nn}{\nonumber}

\newcommand{\be}{\begin{equation}}
\newcommand{\ee}{\end{equation}}
\newcommand{\bear}{\begin{eqnarray}}
\newcommand{\eear}{\end{eqnarray}}

\newcommand{\ket}{\,\rangle}
\newcommand{\bra}{\langle \,}
\newcommand{\eqn}[1]{(\ref{#1})}
\newcommand{\cO}{{\cal O}}

\newcommand{\mL}{\mathcal{L}}

\newcommand{\mF}{\mathcal{F}}

\def\bat{\begin{array}{cc}}

\newcommand{\Frac}[2]{\frac{\displaystyle #1}{\displaystyle #2}}
\newcommand{\Int}{\displaystyle{\int}}
\def\ie{{\it i.e.},\ }
\title{
\vspace*{-2cm}
\flushright {\small IFIC/13-77}
\\
\vspace*{-0.8cm}\flushright {\small FTUAM-13-27}
\\
\vspace*{-0.8cm}\flushright {\small IFT-UAM/CSIC-13-103}
\\
\vspace*{2cm}
Viability of light-Higgs strongly-coupled scenarios
}

\ShortTitle{Viability of light-Higgs strongly-coupled scenarios}

\author{\speaker{Juan Jos\'e Sanz-Cillero}
        \thanks{
    Work  supported in part by the Spanish Government
    and ERDF funds from the European Commission [
    FPA2010-17747, FPA2011-23778, AIC-D-2011-0818, SEV-2012-0249,
    CSD2007-00042], the Generalitat Valenciana
    [PrometeoII/2013/007] and the Comunidad de Madrid [HEPHACOS
    S2009/ESP-1473].
        }
\\
        Departamento de F\'\i sica Te\'orica and Instituto de F\'\i sica Te\'orica, IFT-UAM/CSIC,
        Universidad Aut\'onoma de Madrid, Cantoblanco, E-28049 Madrid, Spain
 \\
        E-mail: \email{juanj.sanz@uam.es}}

\author{Antonio Pich         
\\
        Departament de F\'\i sica Te\`orica, IFIC, Universitat de Val\`encia -- CSIC\\
        Apt. Correus 22085, E-46071 Val\`encia, Spain
 \\
        E-mail: \email{pich@ific.uv.es}}

\author{Ignasi Rosell         
\\
        Departamento de Ciencias F\'\i sicas, Matem\'aticas y de la Computaci\'on, ESET,\\
        Universidad CEU Cardenal Herrera, 
        E-46115 Alfara del Patriarca, Val\`encia, Spain
 \\
        E-mail: \email{rosell@uch.ceu.es}}

\abstract{

Contrary to what is sometimes stated,
the current electroweak precision data easily allow for massive composite resonance states
at the natural EW  scale, \ie well over the TeV.
The oblique parameters $S$ and $T$ are analyzed by means of an effective Lagrangian that
implements the $SU(2)_L\otimes SU(2)_R\to SU(2)_{L+R}$ pattern of electroweak symmetry breaking.
They  are computed  at the one-loop level and
incorporating the newly discovered Higgs-like boson and possible spin--1 composite resonances.
Imposing a proper ultraviolet behaviour is crucial and allows us to determine
$S$ and $T$ at next-to-leading order  in terms of a few resonance parameters.
Electroweak  precision data force the vector and axial-vector states to have
masses above the TeV scale and suggest that  the $W^+W^-$ and $ZZ$ couplings
to  the Higgs-like scalar should be close to the Standard Model value.
Our findings are  generic:  they only rely on symmetry principles and
soft requirements on the short-distance properties of the underlying strongly-coupled theory,
which are widely satisfied in more specific scenarios.

}

\FullConference{The European Physical Society Conference on High Energy Physics -EPS-HEP2013\\
		18-24 July 2013\\
		Stockholm, Sweden}

\begin{document}

\section{Introduction}

In this talk we present the first combined analysis of the oblique parameters
$S$ and $T$~\cite{Peskin:92,Baak:2012kk},
including  the newly discovered Higgs-like boson and possible
spin--1 composite resonances  at the one-loop level~\cite{S+T-letter,S+T-details}.
We consider a general Lagrangian implementing the $SU(2)_L\otimes SU(2)_R\to SU(2)_{L+R}$
pattern of electroweak symmetry breaking (EWSB),
with a non-linear realization of the corresponding
Goldstone bosons~\cite{Appelquist:1980vg}.  
We  consider strongly-coupled models where the gauge symmetry is dynamically
broken by means of some non-perturbative interaction.
Usually, theories of this
kind  do not contain a fundamental Higgs, bringing instead
composite states of different types, in a similar way as it happens in Quantum Chromodynamics.
In the past, electroweak (EW) chiral effective Lagrangians~\cite{Appelquist:1980vg} 
were used for the study of the oblique parameters~\cite{Dobado:1990zh}.
In the recent years, several works have incorporated vector and axial-vector resonances
and performed one-loop computations of $S$ and  $T$ within a similar
$SU(2)_L\otimes SU(2)_R/SU(2)_{L+R}$
effective framework~\cite{S+T-others,S-Orgogozo:11}.
However, they contained unphysical dependences
on the ultraviolet (UV) cut-off, manifesting the need for local contributions to account
for a proper UV completion. Our calculation avoids this problem through the implementation
of short-distance conditions on the relevant Green functions, in order to satisfy the assumed
UV behaviour of the strongly-coupled theory.
As shown in Refs.~\cite{L8+L9,L10}, the dispersive approach   we adopt
avoids all technicalities associated with the renormalization procedure,
allowing for a much more transparent understanding of the underlying physics.

\section{Electroweak effective theory}
\label{sec.lagrangian}

Let us consider a low-energy effective theory containing the Standard Model (SM) gauge bosons coupled
to the EW Goldstones, one scalar state $S_1$ with mass $m_{S_1} = 126$~GeV
and the lightest vector and axial-vector resonance multiplets $V$ and $A$,
which are expected to be the most relevant ones at low energies.
We assume the SM pattern of EWSB and
the scalar field $S_1$ is taken to be a singlet,  
whereas  $V$ and $A$ are introduced as triplets.

The relevant one-loop absorptive diagrams we will compute require  interaction vertices
with at most three legs. In addition, since we just consider contributions from the
lightest channels, $\varphi\varphi$ (two Goldstones)  and $S_1\varphi$ for
the $S$--parameter, and $\varphi B$ and $S_1 B$ for $T$,
we will just need the Lagrangian operators~\cite{S+T-letter,S+T-details}
\bear\label{eq:L_R}
\mL &=&
\bigg( \, 1\, +\,  \Frac{2 \kw }{v}\, S_1\,  \bigg) \, \Frac{v^2}{4} \, \bra  u_\mu u^\mu \ket\,
+ \frac{F_V}{2\sqrt{2}}\, \bra V_{\mu\nu} f^{\mu\nu}_+ \ket
+ \frac{i\, G_V}{2\sqrt{2}}\, \bra  V_{\mu\nu} [u^\mu, u^\nu] \ket
\nn \\
&& \qquad\qquad\qquad \qquad\qquad\qquad
+ \frac{F_A}{2\sqrt{2}}\, \bra A_{\mu\nu} f^{\mu\nu}_- \ket
 + \sqrt{2}\, \lambda_1^{SA}\,  \partial_\mu S_1 \, \bra A^{\mu \nu} u_\nu \ket\, ,
\eear
with $u_\mu =\, - \vec{\sigma}\partial_\mu \vec{\varphi} /v +...$ and the other
chiral tensors are defined in~\cite{S+T-details,S-Higgsless}.
In addition, we will have the Yang-Mills and gauge-fixing terms, with the computation performed
in the Landau gauge.
The term proportional to $\kw$ in Eq.~(\ref{eq:L_R}) contains the coupling of the scalar $S_1$
resonance to two gauge bosons.
For $\kw=1$ one recovers the $S_1 \to\varphi\varphi$ vertex of the SM.
The computation is  performed in the Landau gauge.

\section{Oblique parameters}
\label{sec.observables}

The $S$--parameter  measures the difference between the off-diagonal $W^3 B$ correlator
and its SM value, while $T$ parametrizes the breaking of custodial symmetry~\cite{Peskin:92}:
\begin{eqnarray}
&&S\,=\,  \Frac{16\pi}{g^2}\;\big(e_3 - e_3^{\rm SM}\big)\, ,
\qquad \qquad
T\,=\, \Frac{4\pi  }{g^2   \sin^2{\theta_W}}\; \big(e_1-e_1^{\rm SM}\big)  \,,
\label{eq.S-def}
\end{eqnarray}
with
\vspace*{-0.22cm}
\be
e_3\,=\, \Frac{g}{g'}  \; \widetilde{\Pi}_{30}(0)\, , \qquad \qquad
e_1\,=\, \Frac{1}{M_W^2} ( \Pi_{33}(0) - \Pi_{WW} (0) )\, .
\ee

The tree-level \  Goldstone contribution  \ in $e_3$  \  has been removed from $\Pi_{30}(q^2)$
in the form  ${  \Pi_{30}(q^2)\,=\,q^2\, \widetilde\Pi_{30}(q^2)+g^2 \tan{\theta_W}\, v^2 /4   }$.
For the computation of these oblique parameters  we have made use
of the  dispersive representations~\cite{Peskin:92,S+T-letter,S+T-details}
\bear
S &=& \Frac{16\pi}{g^2 \tan\theta_W}\; \Int_0^\infty \Frac{{\rm dt}}{t}\;
[\rho_S(t)\, -\,\rho_S(t)^{\rm SM} ]\, ,
\label{Peskin-Takeuchi}
\\
T &=& \Frac{4\pi}{ g'^{\, 2} \cos^2\theta_W} \; \Int_0^\infty \Frac{{\rm dt}}{t^2}
\; [\rho_T(t)\, -\, \rho_T(t)^{\rm SM}]\, ,
\label{eq.T-disp-rel}
\eear
with the one-loop spectral functions (we will remain at lowest order in $g$ and $g'$)
\begin{eqnarray}
\rho_S(t) &=&\Frac{1}{\pi}\,\mbox{Im}\widetilde{\Pi}_{30}(t)\, ,
\qquad\qquad
\rho_T(t)  \,=\,
\frac{1}{\pi}\mbox{Im}[\Sigma(t)^{(0)}-\Sigma(t)^{(+)}]\, .
\end{eqnarray}
The first dispersion relation~(\ref{Peskin-Takeuchi})  was  worked out  by
Peskin and Takeuchi~\cite{Peskin:92} and its convergence requires a vanishing
spectral function at short distances.  Since
$\rho_S(t)^{\rm SM}$
vanishes at high energies, the spectral function
$\rho_S(t)$ of the theory we want to analyze must also go to zero for $s\to\infty$.
This  removes from the picture any undesired UV cut-off and  $S$  depends  only
on the physical scales of the problem.
For the computation of $T$, we employ the Ward-Takahashi identity~\cite{Barbieri:1992dq}
which relates the  $\Pi_{33}$ and $\Pi_{WW}$ polarizations with  the EW Goldstone self-energies
$\Sigma^{(0)}$ and $\Sigma^{(+)}$, respectively.
In the Landau gauge one finds  the next-to-leading order (NLO) relation
${ e_1  = \Sigma'(0)^{(0)}\, -\, \Sigma'(0)^{(+)} }$,
with $\Sigma'(t)\equiv\mathrm{\frac{d}{dt}}\Sigma(t)$~\cite{S+T-letter,S+T-details}.
We have computed the  one-loop  contributions to the Goldstone self-energies from
the lightest two-particle absorptive cuts:  $\varphi B$ and $S_1 B$.
Our analysis~\cite{S+T-letter,S+T-details}  shows  that,
once proper short-distance conditions have been imposed on the form-factors that determine $\rho_S(t)$,
the spectral function $\rho_T(t)$ also vanishes at high momentum and
one is allowed to recover  $T$
by means of the UV--converging dispersion relation~\eqn{eq.T-disp-rel}.
Nonetheless, we want to stress that this property, hinted previously by Ref.~\cite{S-Orgogozo:11},
has only been  explicitly checked for the leading channels, $\varphi B$ and $S_1 B$, contributing  to $T$.
The $1/t$  and $1/t^2$  weights in Eqs.~(\ref{Peskin-Takeuchi}) and (\ref{eq.T-disp-rel}),
respectively, enhance the contribution from the lightest thresholds and suppress channels
with heavy states~\cite{L10}. Thus, in this talk we  focus our attention
on the lightest one and two-particle cuts: $\varphi$, $V$, $A$, $\varphi\varphi$ and $S_1\varphi$
for the $S$--parameter;  $\varphi B$ and $S_1 B$ for $T$.
Since the leading-order (LO) determination of $S$ already implies
that the vector and axial-vector masses must be above the TeV scale,
two-particle cuts with $V$ and $A$ resonances are very
suppressed. Their effect was estimated in Ref.~\cite{S-Higgsless} and found to be small.
%

\section{Short-distance constraints: Weinberg sum-rules}


Since we are assuming that weak isospin and parity are good symmetries of the strong
dynamics, the correlator $\Pi_{30}(s)$ can be written in terms of the vector ($R+L$) and axial-vector ($R-L$) two-point functions as \cite{Peskin:92}
\vspace*{-0.25cm}\be
\Pi_{30}(s)\, =\, \frac{g^2 \tan{\theta_W}}{4}\; s\;
\left[ \Pi_{VV}(s) - \Pi_{AA}(s)\right]\, .
\ee
%
%
In asymptotically-free gauge theories the difference $\Pi_{VV}(s)-\Pi_{AA}(s)$ vanishes at $s\to\infty$ as $1/s^3$ \cite{Bernard:1975cd}. This implies two super-convergent sum rules,
known as the 1st and 2nd Weinberg sum-rules (WSRs)~\cite{WSR}.
At LO (tree-level), the 1st and 2nd WSRs imply, respectively,~\cite{Peskin:92,WSR}
\begin{eqnarray}
 F_{V}^2 \,-\, F_{A}^2 \, =\, v^2 \, ,
\label{eq:1stWSR-LO}
\qquad\qquad
F_{V}^2 \, M_{V}^2\, -\, F_{A}^2 \, M_{A}^2  \,=\, 0 \,  ,
\label{eq:2ndWSR-LO}
\end{eqnarray}
where the 1st (2nd) WSR  stems from requiring $\Pi_{VV}(s)-\Pi_{AA}(s)$
to vanish faster than $1/s$  ($1/s^2$)  at short distances.
If both WSRs are valid, one has $M_V < M_A$ and
the vector and axial-vector couplings $F_{V,A}$ can  be determined at LO
in terms of the resonance masses~\cite{Peskin:92,S+T-letter,S+T-details,MHA}.
On the other hand, if only the 1st WSR is assumed then the vector is no longer forced
to be lighter than the axial-vector~\cite{Appelquist:1998xf,Marzocca:2012zn};
all one can say is that $F_V^2>F_A^2$.
It is likely that the  1st WSR is also true in gauge theories
with non-trivial UV fixed points~\cite{S-Orgogozo:11}.
However, the 2nd WSR cannot be used in Conformal Technicolour models \cite{S-Orgogozo:11}
and its validity is questionable in most Walking Technicolour scenarios~\cite{Appelquist:1998xf}.


The $\varphi\varphi$ and $S_1\varphi$ contributions to the spectral function $\rho_S(t)$ are given by
\bear
\rho_S(s)|_{\varphi\varphi}  &=& \theta(s)\;
\Frac{g^2\tan\theta_W}{192\pi^2}\; |\mF^v_{\varphi\varphi}(s)|^2 \, ,
\\
\rho_S(s)|_{S_1\varphi}  &=&\mbox{} -\theta(s-m_{S_1}^2)\;
\Frac{g^2\tan\theta_W}{192\pi^2}\; |\mF^a_{S_1\varphi}(s)|^2
\;\left(1-\Frac{m_{S_1}^2}{s}\right)^3\, ,
\eear
with the   $\varphi\varphi$  and $S_1\varphi$  form-factors, respectively, provided
at LO by~\cite{S+T-letter,S+T-details,L10}
\bear
\mF^v_{\varphi\varphi}(s) &=& 1\, +\, \sigma_V \;\Frac{s}{M_V^2 -s}\, ,
\qquad \qquad
\mF^a_{S_1\varphi}(s)
=
\kw\; \left( \, 1\, +\, \sigma_A \:\Frac{s}{M_A^2 -s} \, \right)\, ,
\eear
with $\sigma_V\equiv F_V G_V/v^2$ and
$\sigma_A\equiv F_A \lambda_1^{\mathrm{SA}}/(\kw v)$.
We will demand these  form factors  to fall as $\cO(1/s)$, \ie
$ \sigma_V=\sigma_A= 1$~\cite{S+T-letter,S+T-details}.
When computing the $T$ parameter at NLO we found that the
$\varphi B$ and $S_1 B$ channels  in the  $\rho_T(t)$   spectral function
were fully determined by the
form-factors $\mF^v_{\varphi\varphi}$
and $\mF^a_{S_1\varphi}$, respectively~\cite{S+T-details}.
This  relation between  the $\varphi\varphi$ vector form-factor
and the $T$--parameter was also previously hinted in Ref.~\cite{S-Orgogozo:11}.
Thus, in addition to making  $\Pi_{30}(t)$ and $\rho_S(t)$  well-behaved  at short distances,
these conditions alone lead to a good high-energy behaviour for  the $\rho_T(t)$
spectral function~\cite{S+T-letter,S+T-details}.

\section{Theoretical predictions at LO and NLO}

At leading order, the tree-level Goldstone self-energies
are identically zero and  one has ${  T_{\rm LO}=0  }$. On the other hand,
for the $S$--parameter one obtains~\cite{Peskin:92,S+T-letter,S+T-details,S-Higgsless}
\bear
S_{\rm LO} &=&  \Frac{4\pi v^2}{M_V^2} \, \bigg(\, 1\, +\, \Frac{M_V^2}{M_A^2}\, \bigg)\,
\qquad\qquad \mbox{(Two WSRs)}\, ,
\label{eq.S-LO-2WSRs}
\\
S_{\rm LO} &> &  \Frac{4\pi v^2}{M_V^2}
\qquad \qquad\qquad\qquad \mbox{(Only the 1st WSRs; $M_V<M_A$)}\, ,
\label{eq.S-LO-1WSR}
\eear
with the last inequality flipping sign (becoming an identity) in the
inverted-mass scenario $M_V>M_A$~\cite{Appelquist:1998xf,Marzocca:2012zn}
(degenerate-mass scenario $M_V=M_A$).
Eq.~(\ref{eq.S-LO-2WSRs}) assumes the validity of the two WSRs,
while only the 1st WSR is taken into account in Eq.~(\ref{eq.S-LO-1WSR}),
but assuming $M_V < M_A$. In both cases,
the resonance masses need to be heavy enough to comply  with the stringent experimental limits
on $S$~\cite{Baak:2012kk}, implying    $M_V>1.5$~TeV (2.3~TeV) at the 3$\sigma$ (1$\sigma$) level.

At NLO, the requirement that Im$\widetilde{\Pi}_{30}(s)$ vanishes at short distances allows us
to reconstruct the full correlator $\Pi_{30}(s)$ through a one subtracted dispersion
relation~\cite{S+T-letter,S+T-details,L10,S-Higgsless}:
\be
\left. \Pi_{30}(s) \right|_{\mathrm{NLO}}  \; =\;
\frac{g^2\tan{\theta_W} }{4}\;  s \;  \left(\frac{v^2}{s}+  \frac{F_{V}^{r\,2}}{M_{V}^{r\,2}-s}
- \frac{F_{A}^{r\,2}}{M_{A}^{r\,2}-s} \; +\; \overline{\Pi}(s)\right)\, ,
\label{eq.T-NLO}
\ee
with  the renormalized $F_R^r$ and $M_R^r$ and
the finite one-loop contribution  $\overline{\Pi}(s)$, fully  determined
by Im$\widetilde{\Pi}_{30}(s)$  (see App.~A of Ref.~\cite{S-Higgsless}).
By imposing the WSRs at NLO, one obtains NLO conditions on the high-energy expansion
of $\Pi_{30}(s)|_{\rm NLO}$ in powers of $1/s$.
Its real  and imaginary parts allow us to
constrain the renormalized resonance couplings $F_{V,A}^{r\, \, 2}$ and
produces the condition $\kw = M_V^2/M_A^2$
(in the case with two WSRs), respectively.
Thus, for the NLO $S$--parameter one finds~\cite{S+T-letter,S+T-details}
\bear
S & =&  4 \pi v^2 \left(\frac{1}{M_{V}^2}+\frac{1}{M_{A}^2}\right)
\, + \,  \frac{1}{12\pi}\,
\left[ \log\frac{M_V^2}{m_{H}^2}  -\frac{11}{6}
\right.
\left.+\;\frac{M_V^2}{M_A^2}\log\frac{M_A^2}{M_V^2}
 - \frac{M_V^4}{M_A^4}\, \bigg(\log\frac{M_A^2}{m_{S_1}^2}-\frac{11}{6}\bigg) \right]
\nn \\
&&\quad\qquad \qquad\qquad
\qquad \qquad\qquad \qquad\qquad \qquad\qquad \qquad \qquad\qquad \mbox{(Two WSRs)}\, ,
\label{eq.1+2WSR}
\\
S &> &  \Frac{4 \pi v^2}{M_{V}^{2}}
 \,\,+\,\,
\Frac{1}{12\pi}
\left[ \left(\ln\Frac{M_V^2}{m_{H}^2}-\Frac{11}{6}\right)
 - \,\kw^2\, \left(\log\Frac{M_A^2}{m_{S_1}^2}-\Frac{17}{6}
 + \Frac{M_A^2}{M_V^2}\right) \right]
\nn\\
&&\qquad\qquad
\qquad \qquad\qquad \qquad\qquad \qquad
 \qquad\qquad \mbox{(Only the 1st WSR; $M_V<M_A$)}\, ,
\label{eq.S-NLO-1WSR}
\eear
where $m_H$ sets the reference Higgs mass in the definition of the oblique parameters.
We have used the renormalized masses in the NLO expressions and the superscript $r$
is dropped from now on.
As in the LO case, in the case $M_V>M_A$~\cite{Appelquist:1998xf,Marzocca:2012zn}
($M_A=M_V $), the inequality~(\ref{eq.S-NLO-1WSR})
flips direction (becomes an identity).

As we saw in the previous section,
one also has  $\rho_T(t)\stackrel{t\to \infty}{\longrightarrow} 0$ for the
$\varphi B$ and $S_1 B$ channels
once the $\rho_S(t)$ spectral function constraints  $\sigma_V=\sigma_A=1$ are imposed
and the  form-factors vanish at high energies.
The $T$ dispersion relation~(\ref{eq.T-disp-rel}) becomes then  UV convergent and
yields~\cite{S+T-letter,S+T-details}
\begin{equation}
 T\; =\;  \frac{3}{16\pi \cos^2 \theta_W}\; \left[ 1 + \log{\frac{m_{H}^2}{M_V^2}}
 - \kw^2\, \left( 1 + \log{\frac{m_{S_1}^2}{M_A^2}} \right)  \right] \, .
\label{eq:T}
\end{equation}
Terms of $\mathcal{O}(m_{S_1}^2/M_{A}^2)$ have been neglected
in Eqs.~(\ref{eq.1+2WSR})--(\ref{eq:T}). After imposing the high-energy constraints,
the $S$ and $T$ determinations can be written in terms of two (three)  parameters,
{\it e.g.}, $M_V$ and $\kw$  ($M_V$, $M_A$ and $\kw$), in the case with two WSRs
(with only the 1st WSR).

\section{Phenomenology}
\label{sec.conclusions}

\begin{itemize}

\item[]{\bf 1)  Case with two WSRs:}
In the more restrictive scenario,  we find at 68\% (95\%) CL  (Fig.~\ref{fig.2WSR}):
\be
0.97 \; (0.94)\; <\; \kw\; <\; 1\, ,
\qquad\qquad
M_A\; >\; M_V\; >\; 5\; (4)\: \mathrm{TeV}.
\ee
As $\kw= M_V^2/M_A^2$ due to the 2nd WSR at NLO,
the vector and axial-vector turn out to be  quite degenerate.

\begin{figure}
\begin{center}
\includegraphics[scale=0.53]{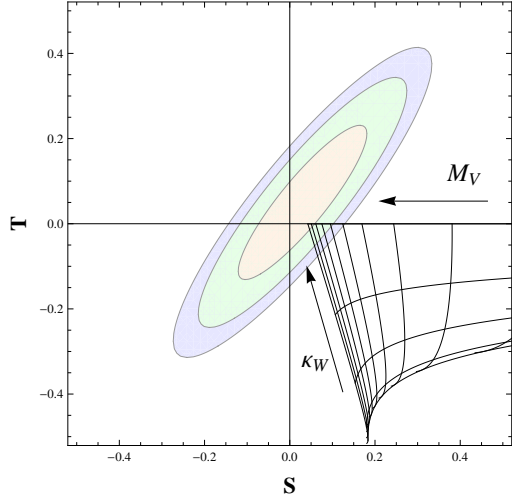}
\caption{\small{NLO determinations of $S$ and $T$, imposing the two WSRs.
The grid lines correspond to $M_V$ values from $1.5$ to $6.0$~TeV, at intervals of $0.5$~TeV, and $\kw= 0, 0.25, 0.50, 0.75, 1$.
The arrows indicate the directions of growing  $M_V$ and $\kw$.
The ellipses give the experimentally allowed regions
at 68\% (orange), 95\% (green) and 99\% (blue) CL~\cite{Baak:2012kk}.}}
\label{fig.2WSR}
\end{center}
\end{figure}

\item[]{\bf 2)  Case with only the 1st WSR:}
The previous stringent bounds get softened when only the 1st WSR is required to be valid.
On general grounds, one would expect this scenario to satisfy the mass hierarchy $M_V<M_A$.
Assuming a moderate splitting $0.5<M_V/M_A<1$, we obtain (68\% CL)
\be
0.84 \; <\;\kw\; <\; 1.3\, ,
\qquad\qquad
M_V\; >\; 1.5\; \mathrm{TeV}.
\ee
Slightly larger departures from the SM can be achieved by considering a larger mass splitting.
\\
When the resonance masses become degenerate, the allowed range for the scalar coupling
shrinks to $0.97 < \kw < 1.3$ (68\% CL) (black band Fig.~\ref{fig.1WSR}, right-hand side).
A heavier resonance mass is also necessary,
with $M_V=M_A > 1.8$~TeV (68\% CL).
\\
Finally, in the inverted-mass scenario,   we obtain  the upper bound
$\kw<2$ (68\% CL) for  $1<M_V/M_A<2$.
%
%
Nonetheless,  if no vector resonance is seen below the TeV ($M_V>1$~TeV)
the scalar coupling  becomes again constrained to be around $\kw \simeq 1$ for $1<M_V/M_A<2$,
with the 68\% CL interval $0.7 < \kw < 1.9$.
The outcomes for various mass splittings in the different scenarios with only the 1st WSR
(normal-ordered, degenerate and inverted-mass) can be observed in Fig.~\ref{fig.1WSR}.

\end{itemize}
In summary, contrary to what is sometimes stated,
the current electroweak precision data easily allow for resonance states
at the natural EW  scale, \ie well over the TeV.
The present results are in good agreement with the $H\to WW,ZZ$ couplings measured at LHC,
compatible with  the Standard Model up to deviations of
the order of 20\% or smaller~\cite{Aad:2013wqa+Chatrchyan:2013lba}).
These conclusions are generic, since we have only used mild assumptions about the
UV behavior of the underlying strongly-coupled theory,
and can be easily particularized to more specific models obeying the
$SU(2)_L\otimes SU(2)_R\to SU(2)_{L+R}$ EWSB pattern.

\begin{figure}
\begin{center}
\includegraphics[scale=0.5]{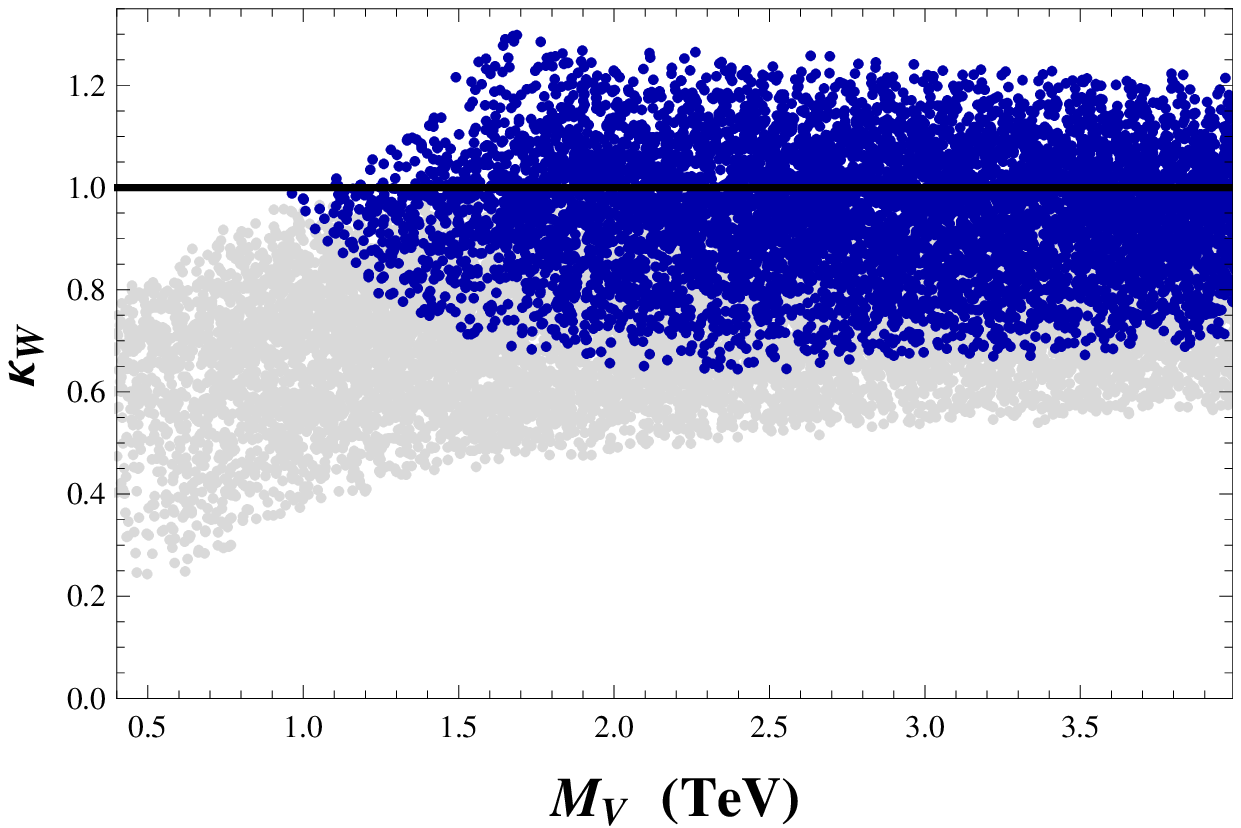}
\hspace*{1cm}
\includegraphics[scale=0.5]{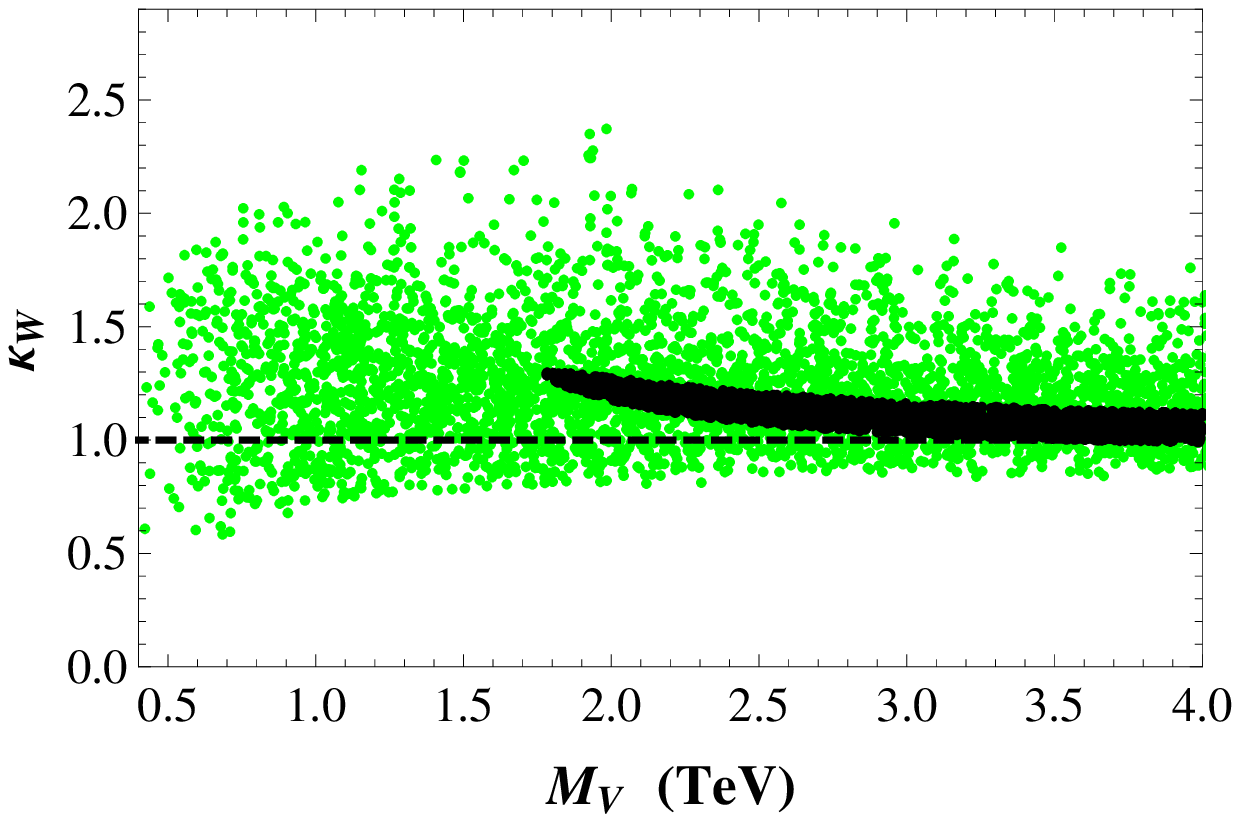}
\caption{\small {\bf
Left-hand side:}
Scatter plot for the 68\% CL region,
in the case when only the 1st WSR is assumed, for $M_V<M_A$.
The dark blue and light gray regions correspond, respectively,
to $0.2<M_V/M_A<1$ and $0.02<M_V/M_A<0.2$.
%
{\bf Right-hand side:}
68\% CL region  with only the 1st WSR  for  the degenerate  and inverted-hierarchy  scenarios.
The black (dark) and green (lighter) regions
correspond, respectively,  to
$M_V=M_A$ and $1<M_V/M_A<5$.
We consider $M_{V,A}>0.4$~TeV in both plots.
}
\label{fig.1WSR}
\end{center}
\end{figure}


\begin{thebibliography}{99}

\bibitem{Peskin:92}
  M. E. Peskin and T. Takeuchi,
  Phys.\ Rev.\ D {\bf 46} (1992) 381;
  Phys.\ Rev.\ Lett.\  {\bf 65} (1990) 964.

\bibitem{Baak:2012kk} M.~Baak {\it et al.},
  Eur.\ Phys.\ J.\ C {\bf 72} (2012) 2205;   
%
http://gfitter.desy.de/ .



\bibitem{S+T-letter}
  A.~Pich, I.~Rosell and J.~J.~Sanz-Cillero,
Phys.\ Rev.\ Lett.\  {\bf 110} (2013) 181801;   
    [arXiv:1307.1958~[hep-ph]].


\bibitem{S+T-details}
    A. Pich, I. Rosell and J.J. Sanz-Cillero,
    [arXiv:1310.3121 [hep-ph]].



\bibitem{Appelquist:1980vg}
  T.~Appelquist and C.~W.~Bernard,
  Phys.\ Rev.\ D {\bf 22} (1980) 200;
%
%
%
  A.~C.~Longhitano,
  Phys.\ Rev.\ D {\bf 22} (1980) 1166;
%
  Nucl.\ Phys.\ B {\bf 188} (1981) 118.



\bibitem{Dobado:1990zh}
  A.~Dobado, D.~Espriu and M.~J.~Herrero,
  Phys.\ Lett.\ B {\bf 255} (1991) 405.



\bibitem{S+T-others}
%
  S.~Matsuzaki {\it et al.},     
  Phys.\ Rev.\ D {\bf 75} (2007) 073002;
 075012;
%
    R. Barbieri {\it et al.},     
    Phys.\ Rev.\ D {\bf 78} (2008) 036012;
%
    O. Cat\`a and J.F. Kamenik,
    Phys.\ Rev.\ D {\bf 83} (2011) 053010;
%
%
  A.~Orgogozo and S.~Rychkov,
  JHEP {\bf 1306} (2013) 014.



%
\bibitem{S-Orgogozo:11}
  A.~Orgogozo and S.~Rychkov,
  JHEP {\bf 1203} (2012) 046;




\bibitem{L8+L9}
   A.~Pich, I.~Rosell and J.J.~Sanz-Cillero,
  JHEP {\bf 0701} (2007) 039;
%
  JHEP {\bf 1102} (2011) 109.




\bibitem{L10}
%
  A.~Pich, I.~Rosell and J.~J.~Sanz-Cillero,
  JHEP {\bf 0807} (2008) 014.




\bibitem{S-Higgsless}
  A.~Pich, I.~Rosell and J.~J.~Sanz-Cillero,
  JHEP {\bf 1208} (2012) 106.  



\bibitem{Barbieri:1992dq}
  R.~Barbieri {\it et al.},
  Nucl.\ Phys.\ B {\bf 409} (1993) 105.



\bibitem{Bernard:1975cd}
  C.~W.~Bernard {\it et al.},       
  Phys.\ Rev.\ D {\bf 12} (1975) 792.

\bibitem{WSR}
  S.~Weinberg,
  Phys.\ Rev.\ Lett.\  {\bf 18} (1967) 507.


\bibitem{MHA}
    M. Knecht and E. de Rafael,
    Phys. Lett. B {\bf 424}  (1998) 335.   


\bibitem{Appelquist:1998xf}
  T.~Appelquist and F.~Sannino,
  Phys.\ Rev.\ D {\bf 59} (1999) 067702.   

\bibitem{Marzocca:2012zn}
    D. Marzocca, M. Serone and  J.  Shu,
    JHEP {\bf 1208} (2012) 013
    [arXiv:1205.0770 [hep-ph]].



\bibitem{Aad:2013wqa+Chatrchyan:2013lba}
  ATLAS Collaboration,
  Phys.\ Lett.\ B {\bf 726} (2013) 88;  
%
  ATLAS-CONF-2013-079 (July 19, 2013);
%
%
  CMS Collaboration,
  JHEP {\bf 06} (2013) 081;     
%
  CMS-PAS-HIG-13-005 (April 17, 2013).





\end{thebibliography}
\end{document}